\documentclass[twocolumn,showpacs,aps,prd,superscriptaddress]{revtex4}

\usepackage{graphicx}
\usepackage{dcolumn}
\usepackage{amsmath}
\usepackage{units}
\usepackage{pstricks}

\input{pubboard/babarsym}

\newcommand{\Btag}{\ensuremath{\B_\mathrm{tag}}}
\newcommand{\Bztojpsiks}{\ensuremath{\Bz\to\jpsi\KS}}
\newcommand{\Bztophiks} {\ensuremath{\Bz\to\phi\KS}}
\newcommand{\Bztoetapks} {\ensuremath{\Bz\to\eta'\KS}}

\newcommand{\Bztokpkmks} {\ensuremath{\Bz\to K^+ K^- \KS}}
\newcommand{\Bztokspiz} {\ensuremath{\Bz\to\KS\piz}}
\newcommand{\Bztokzpiz} {\ensuremath{\Bz\to K^{0}\piz}}
\newcommand{\ckspiz} {\ensuremath{C_{\KS\piz}}}
\newcommand{\skspiz} {\ensuremath{S_{\KS\piz}}}
\def\cf {\ensuremath{C_f}}
\def\sf {\ensuremath{S_f}}

\newcommand{\Brec}{\ensuremath{B_{\CP}}}
\newcommand{\Ups}{\ensuremath{\Upsilon}}
\newcommand{\dte}{\ensuremath{\sigma(\deltat)}}
\newcommand{\thetacms}{\ensuremath{\theta_{B}^*}}
\newcommand{\costhetacms}{\ensuremath{\cos\thetacms}}
\newcommand{\mmiss}{\ensuremath{m_\text{miss}}}
\newcommand{\mb}{\ensuremath{m_{B}}}
\newcommand{\rhom}{\ensuremath{\rho^{-}}}

\newcommand{\BABARPubYear}    {05}
\newcommand{\BABARPubNumber}  {01}
\newcommand{\SLACPubNumber} {11048}

\def\figurebox#1#2#3{%
    \def\arg{#3}%
    \ifx\arg\empty
    {\hfill\vbox{\hsize#2\hrule\hbox to #2{\vrule\hfill\vbox to #1{\hsize#2\vfill}\vrule}\hrule}\hfil
l}%
    \else
    {\hfill\epsfbox{#3}\hfill}%
    \fi}

\long\def\inst#1{\par\nobreak\kern 4pt\nobreak
    {\it #1}\par\vskip 10pt plus 3pt minus 3pt}

\begin{document}

\preprint{\babar-PUB-\BABARPubYear/\BABARPubNumber}
\preprint{SLAC-PUB-\SLACPubNumber}

\begin{flushleft}
  \babar-PUB-\BABARPubYear/\BABARPubNumber\\
  SLAC-PUB-\SLACPubNumber\\
\end{flushleft}

\title{ {\Large \bf \boldmath Measurement of the Branching Fraction and
    the \CP-Violating Asymmetry for the Decay \Bztokspiz{}}  }

%
\author{B.~Aubert}
\author{R.~Barate}
\author{D.~Boutigny}
\author{F.~Couderc}
\author{Y.~Karyotakis}
\author{J.~P.~Lees}
\author{V.~Poireau}
\author{V.~Tisserand}
\author{A.~Zghiche}
\affiliation{Laboratoire de Physique des Particules, F-74941 Annecy-le-Vieux, France }
\author{E.~Grauges-Pous}
\affiliation{IFAE, Universitat Autonoma de Barcelona, E-08193 Bellaterra, Barcelona, Spain }
% ...
\collaboration{The \babar\ Collaboration}
\noaffiliation

\date{\today}

\begin{abstract}
  We measure the branching fraction and the \CP-violating asymmetry of
  \Bztokspiz{} decays with $227$ million $\Y4S\to\BB$ events collected
  with the \babar{} detector at the \pep2{} asymmetric-energy \epem{}
  collider at SLAC. We obtain a branching fraction $\BR{}(\Bztokzpiz)
  = (11.4 \pm 0.9 \pm 0.6) \times 10^{-6}$ and \CP-violating asymmetry
  parameters $\ckspiz = 0.06 \pm 0.18 \pm 0.03$ and $\skspiz =
  0.35^{+0.30}_{-0.33} \pm 0.04$, where the first error is statistical
  and the second systematic.
\end{abstract}


\maketitle

\CP{} violation effects in decays of $B$ mesons that are dominated by
penguin $\b\to\s\qbar\q$ ($\q=\u,\d,\s$) transitions are
potentially sensitive to contributions from physics beyond the
Standard Model (SM)~\cite{ref:newphysics}. The \B{}-factory
experiments have explored time-dependent \CP{}-violating (CPV)
asymmetries in several such decays~\cite{ref:cc}, namely
\Bztophiks{}~\cite{Abe:2003yt,Aubert:2004ii},
\Bztoetapks{}~\cite{Abe:2003yt,Aubert:2003bq},
\Bztokpkmks{}~\cite{Abe:2003yt,Aubert:2004ta}, $\Bz\to
f_{0}\KS$~\cite{Aubert:2004am} and \Bztokspiz{}~\cite{Aubert:2004xf}.
Within the SM these asymmetries are expected to be consistent with the
measurement of \stwob{} in charmonium modes originating from the
tree-level $\b\to\c\cbar\s$ transition. These comparisons must take
into account possible deviations for each mode, within the SM, due to
contributions of other diagrams with different phases and rescattering
effects. At this point none of the modes above shows a significant
deviation from the SM expectation~\cite{Ligeti:2004ak}.  A major goal
of the $B$-factory experiments is to reduce the experimental
uncertainties of these measurements in order to improve the sensitivity
to beyond-the-Standard-Model effects.

In this letter we present improved measurements of the CPV asymmetry
in the decay \Bztokspiz{}, using data collected with the \babar{}
detector at the \pep2{} asymmetric-energy \epem{} collider, 
amounting to $226.6\pm2.5$ million $\Y4S\to\BB$ decays.  In
the SM this decay is dominated by a top-quark-mediated
$\b\to\s\dbar\d$ penguin amplitude. If other contributions, such as
the CKM suppressed $\b\to\s\ubar\u$ tree amplitude, are ignored, the
CPV asymmetry is governed by \stwob{}~\cite{Fleischer:1995cg}, where
$\beta\equiv\arg[-V_{cd}V_{cb}^{*}/V_{td}V_{tb}^{*}]$ and $V$ is the
Cabibbo-Kobayashi-Maskawa (CKM) quark mixing matrix~\cite{CKM}.
The bound on the deviation from \stwob{} due to SM contributions with
a different weak phase is about $0.2$ from SU(3) flavor
symmetry~\cite{Gronau:2003kx} and about $0.1$ in model-dependent QCD
calculations~\cite{ref-qcdbounds}. We also present an update of our
measurement of the branching fraction of
\Bztokzpiz{}~\cite{Aubert:2003sg}, which, when combined with
measurements of other $B \to K \pi$ branching fractions, can be used
to extract the CKM angle
$\gamma\equiv\arg[-V_{ud}V_{ub}^{*}/V_{cd}V_{cb}^{*}]$~\cite{Gronau:1994bn}.

The \babar{} detector, fully described in~\cite{ref:babar}, provides
charged particle tracking through a combination of a five-layer
double-sided silicon micro-strip detector (SVT) and a 40-layer drift
chamber (DCH), both operating in a \unit[1.5]{T} magnetic field.
Charged-kaon and -pion identification is achieved through measurements
of specific energy-loss ($dE/dx$) in the tracking system and of the
Cherenkov angle ($\theta_c$) in a detector of internally reflected
Cherenkov light (DIRC).  A CsI(Tl) electromagnetic calorimeter (EMC)
provides photon detection and electron identification.  Finally, the
instrumented flux return (IFR) of the magnet allows discrimination of
muons from pions. For event simulation we use the Monte Carlo event
generator EVTGEN~\cite{Lange:2001uf} and
GEANT4~\cite{Agostinelli:2002hh}.

At the \Y4S{} resonance time-dependent CPV asymmetries are extracted
from the distribution of the difference of the proper decay times,
$\deltat\equiv t_{\CP}-t_\text{tag}$, where $t_{\CP}$ refers to the
decay time of the signal \B{} (\Brec{}) and $t_\text{tag}$ to that of
the other \B{} (\Btag). The \deltat{} distribution for $\Brec\to f$
follows
\begin{eqnarray}
  \label{eqn:td}
  \lefteqn{{\cal P}_{\pm}(\deltat) \; = \; \frac{e^{-|\deltat|/\tau}}{4\tau} \times }\; \\
   && \left[ \: 1 \; \pm \;
    \: S_f \sin{( \deltamd\deltat)} \mp C_f \cos{( \deltamd\deltat)} \: \right] \; , \nonumber
\end{eqnarray}
where the upper (lower) sign corresponds to \Btag{} decaying as \Bz{}
(\Bzb), $\tau$ is the \Bz{} lifetime and \deltamd{} is the mixing
frequency. The coefficients $C_f$ and $S_f$ can be expressed in terms
of the \Bz-\Bzb{} mixing amplitude and the decay amplitudes for
$\Bz\to f$ and $\Bzb\to f$~\cite{lambda}. For decays to a \CP{}
eigenstate, like $\KS\piz$, $C_f$ vanishes unless there is direct
\CP{} violation. If \Bztokspiz{} proceeds purely through a top-quark
penguin, $C_{\KS\piz}=0$ and $S_{\KS\piz}=\sin(2\beta + 2 \beta_s)$,
where $\beta_s\equiv \arg [ - V_{ts}V_{tb}^{*}/V_{cs} V_{cb}^{*}]$ is
small.

We search for \Bztokspiz{} decays in \BB{} candidate events
selected using charged-particle multiplicity and event
topology~\cite{ref:Sin2betaPRD}.  We reconstruct $\KS\to\pip\pim$
candidates from pairs of oppositely charged tracks.  The two-track
combinations must form a vertex with a $\chi^2$ consistency greater
than $0.001$ and a $\pip\pim$ invariant mass within
\unit[$11.2$]{\mevcc} of the nominal \KS\ 
mass~\cite{Eidelman:pdg2004}.  We form $\piz\to\gamma\gamma$
candidates from pairs of photon candidates in the EMC, each of which
is isolated from any charged tracks, carries a minimum energy of
\unit[$50$]{\mev}, and has the expected lateral shower shape.
Candidates for \Bztokspiz{} are formed from $\KS\piz$ combinations
and constrained to originate from the \epem{} interaction point using
a geometric fit. We require that the consistency of the $\chi^2$ of
the fit, which has one degree of freedom, be greater than $0.001$.  We
extract the \KS{} decay length $L_{\KS}$ and the $\piz\to\gamma\gamma$
invariant mass from this fit and require \unit[$110 < m_{\gamma\gamma}
< 160$]{\mevcc} and $L_{\KS}$ greater than $5$ times its uncertainty.

For each $B$ candidate we compute two kinematic variables, namely the
invariant mass \mb{} and the missing mass $\mmiss = \sqrt{(q_{\epem} -
\tilde{q}_B)^2}$, where $q_{\epem}$ is the four-momentum of the initial
\epem{} system and $\tilde{q}_B$ is the four-momentum of the
\Bztokspiz{} candidate after a mass constraint on the \Bz{} is
applied. By construction the linear correlation coefficient between
\mmiss{} and \mb{} vanishes.  Compared to the kinematic variables
$\DeltaE = E^{*}_B - \frac{1}{2}\sqrt{s}$ and $\mes =
\sqrt{\frac{1}{4} s - p^{*2}_B}$ (where $s=q_{\epem}^2$ and the
asterisk denotes the \epem{} rest frame), which were used in our
previous analysis of this mode~\cite{Aubert:2004xf}, the present
combination of variables leads to a smaller correlation and a better
background suppression for modes containing a high-momentum \piz{} or
photon. From simulation studies we determine the signal resolution for
\mb{} to be about \unit[$40$]{\mevcc}. The distribution exhibits a
low-side tail from energy leakage out of the EMC. The signal
resolution for \mmiss{}, about \unit[$5$]{\mevcc}, is dominated by the
beam-energy spread.  We select candidates with \mb{} within
\unit[$150$]{\mevcc} of the nominal \Bz{} mass~\cite{Eidelman:pdg2004}
and with \unit[$5.11<\mmiss<5.31$]{\gevcc}. The region
\unit[$\mmiss<5.2$]{\gevcc} is devoid of signal and used for
background characterization.

To suppress background from continuum $\epem\to\qqbar$
($\q=\u,\d,\s,\c$) events, we exploit differences in both production
and decay properties. We require $|\costhetacms|<0.9$, where
$\thetacms$ is the angle between the \B{}-candidate momentum and the
$e^{-}$ momentum in the \epem{} rest frame. For true \B{} mesons the
distribution of $\costhetacms$ is proportional to
$1 - \cos^2\!\thetacms$, whereas for continuum events it is nearly
flat. To exploit the jet-like topology of continuum events, we
calculate the ratio $L_{2}/L_{0}$ of two Legendre moments defined as
$L_j\equiv\sum_i |{\bf p}^*_i| |\cos \theta^*_i|^j$, where ${\bf
  p}^*_i$ is the momentum of particle $i$ in the \epem{} rest frame,
$\theta^*_i$ is the angle between ${\bf p}^*_i$ and the thrust axis of
the \B{} candidate and the sum runs over all reconstructed particles
except for the \B{}-candidate daughters.  We require $L_2/L_0<0.55$, which
suppresses the background by more than a factor $3$ at the cost of
approximately \unit[$10$]{\%} loss in signal efficiency.  After all
selections are applied the average candidate multiplicity in events
with at least one candidate is approximately $1.007$.  When there 
are multiple candidates, we select the candidate with a reconstructed
\piz{} mass closest to the expected value.

For each \Bztokspiz{} candidate we examine the remaining tracks 
in the event to determine the decay vertex position
and the flavor of \Btag. Using a neural network based on kinematic and particle
identification information~\cite{ref:sin2betaPRL02} each event is
assigned to one of five mutually exclusive tagging categories,
designed to combine flavor tags with similar performance and \deltat{}
resolution.  We parameterize the performance of this algorithm in a
data sample ($B_{\rm flav}$) of fully reconstructed $\Bz\to
D^{(*)-}\pip/\rho^+/a_1^+$ decays. The average effective tagging
efficiency obtained from this sample is $Q = \sum_c \epsilon_S^c (1-2w^c)^2=0.299\pm 0.005$,
where $\epsilon_S^c$ and $w^c$ are the efficiencies and mistag
probabilities, respectively, for events tagged in category
$c=1,2,\cdots5$. For the background, the fraction of events
($\epsilon_B^c$) and the asymmetry in the rate of $\Bz$ versus $\Bzb$
tags in each tagging category are extracted from a fit to the data.

The proper-time difference is extracted from the separation of the
\Brec{} and \Btag{} decay vertices. The \Btag{} vertex is
reconstructed inclusively from the remaining charged particles in the
event~\cite{ref:Sin2betaPRD}. To reconstruct the \Brec{} vertex from
the single \KS{} trajectory we exploit the knowledge of the average
interaction point (IP), which is determined on a run-by-run basis from
the spatial distribution of vertices from two-track events.  We
compute \deltat{} and its uncertainty from a geometric fit to the
$\Ups(4S)\to\Bz\Bzb$ system that takes this IP constraint into
account. We further improve the sensitivity to \deltat{} by
constraining the sum of the two $B$ decay times
($t_{\CP}+t_\text{tag}$) to be equal to $2\:\tau_{\Bz}$ with an
uncertainty $\sqrt{2}\; \tau_{\Bz}$, which effectively constrains the
two vertices to be near the \Y4S{} line of flight. We have verified in
a Monte Carlo simulation that this procedure provides an unbiased
estimate of \deltat{}.

The per-event estimate of the uncertainty on \deltat{} reflects the
strong dependence of the \deltat{} resolution on the $\KS$ flight
direction and on the number of SVT layers traversed by the $\KS$ decay
daughters. In about \unit[$60$]{\%} of the events both pion tracks 
are reconstructed from at least 4 SVT hits, leading to sufficient resolution
for the time-dependent measurement. The average \deltat{} resolution
in these events is about \unit[$1.0$]{ps}. For events which fail this
criterion or for which \unit[$\dte>2.5$]{ps} or
\unit[$\deltat>20$]{ps}, the \deltat{} information is not used.
However, since \cf{} can also be extracted from flavor tagging
information alone, these events still contribute to the measurement of
\cf{}.

We extract the signal yield, \sf{} and \cf{} from an unbinned
maximum-likelihood fit to $m_B$, \mmiss{}, $L_{2}/L_{0}$,
\costhetacms{}, \deltat{} and the flavor tag variables. By exploiting
sideband regions in data for the background and simulated events
for the signal, we have verified that with the selection presented
above these observables are sufficiently independent that we can
construct the likelihood from the product of one-dimensional
probability density functions (PDFs).  The PDFs for signal events are
parameterized from simulated events or from the $B_{\rm flav}$ sample.
For background PDFs we select a functional form that describes the
data in the sideband regions of the other observables, in which
backgrounds dominate.  We include these regions in the fitted sample
and simultaneously extract the parameters of the background PDFs along
with the signal yield and CPV asymmetries.

We obtain the PDF for the \deltat{} of signal events from the
convolution of Eq.~\ref{eqn:td} with a resolution function ${\cal
  R}(\delta t \equiv \deltat -\deltat_{\rm true},\sigma_{\deltat})$.
The resolution function is parameterized as the sum of two Gaussians
with a width proportional to the reconstructed $\sigma_{\deltat}$, and
a third Gaussian with a fixed width of
\unit[$8$]{ps}~\cite{ref:Sin2betaPRD}. The first two Gaussians have a
non-zero mean, proportional to $\sigma_{\deltat}$, to account for the
small bias in \deltat{} from charm decays on the \Btag{} side.  We
have verified in simulation that the parameters of ${\cal R}(\delta t,
\sigma_{\deltat})$ for \Bztokspiz{} events are similar to those
obtained from the $B_{\rm flav}$ sample, even though the distributions
of $\sigma_{\deltat}$ differ considerably. We therefore extract these
parameters from a fit to the $B_{\rm flav}$ sample. We assume that the
background consists of prompt decays only and find that the \deltat{}
distribution is well described by a resolution function with the same
functional form as used for signal events. The parameters of the
background function are determined in the fit.

To extract the yield and the CPV asymmetries we maximize the logarithm
of the extended likelihood 
{\small\begin{eqnarray*} 
  \lefteqn{{\cal L}(\sf,\cf,N_S,N_B,f_S,f_B,\vec{\alpha}) \; = \;
    e^{-(N_S+N_B)} \; \times } \\
  & & \prod_{i \in I} \left[ N_S f_S
      \epsilon^{c}_S{\cal P}_S(\vec{x}_i,\vec{y}_i;\sf,\cf) +
      N_B f_B \epsilon^{c}_B {\cal P}_B(\vec{x}_i,\vec{y}_i;\vec{\alpha}) \right] \times \\
 & & \prod_{i \in II} \left[ N_S (1-f_S)
    \epsilon^{c}_S {\cal P}'_S(\vec{x}_i;\cf) + N_B (1-f_B)
    \epsilon_B^{c} {\cal P}'_B(\vec{x}_i;\vec{\alpha}) \right],
\end{eqnarray*}}
\noindent where $I$ ($II$) is the subset of events with (without) \deltat{}
information. The probabilities ${\cal P}_S$ (${\cal P}'_S$) and ${\cal
  P}_B$ (${\cal P}'_b$) are products of PDFs for signal ($S$) and
background ($B$) hypotheses evaluated for the measurements
$\vec{x}_i=\{\mb,\mmiss,L_{2}/L_{0},\costhetacms,\text{tag},\text{tagging
  category}\}$ and $\vec{y}_i=\{\deltat,\sigma_{\deltat}\}$. Along
with the signal yield $N_S$ and the coefficients \sf{} and \cf{}, the
fit extracts the background yields $N_B$, the fractions of events with
$\deltat$ information $f_S$ and $f_B$, and the remaining parameters,
collectively denoted by $\vec{\alpha}$. These include all parameters
of background PDFs and some parameters of the signal PDFs, such as the
mean values of \mb{} and \mmiss{}.


Fitting the data sample of $9726$ \Bztokspiz{} candidates, we find
\mbox{$N_S=300\pm 23$} signal decays with 
\(
  \skspiz \; = \; 0.35 \:^{+0.30}_{-0.33} \: \text{(stat)} \: \pm \: 0.04 \: \text{(syst)}
\)
and
\(
  \ckspiz \; = \; 0.06\: \pm \: 0.18 \: \text{(stat)} \: \pm 0.03 \: \text{(syst)}.
\)
The number of signal decays with \deltat{} information is $f_s N_s=186\pm18$.
The total detection efficiency for \Bztokspiz{} decays with
$\KS\to\pip\pim$ and $\piz\to\gamma\gamma$ is
\unit[($34.1\pm1.8$)]{\%}. With the \KS{} and \piz{} branching
fractions taken from~\cite{Eidelman:pdg2004} and assuming equal
production of charged and neutral $B$ mesons at the \Y4S{} resonance,
we obtain a branching fraction 
\(
  \BR{}(\Bztokzpiz) \; = \; (11.4 \: \pm \: 0.9 \: \text{(stat)} \: \pm \: 0.6
  \: \text{(syst)} \: ) \; \times \; 10^{-6} \; .
\)
The evaluation of the systematic uncertainties is described below.

\begin{figure}[!tbp]
  \includegraphics[width=\linewidth]{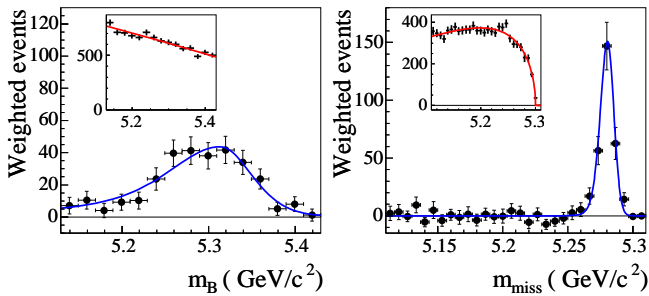}
  \caption{Signal and background (inset) distributions, obtained with the
    weighting technique described in the text, for \mb{} (left) and
    \mmiss{} (right). The curves represent the PDFs used in the fit
    and are normalized to the fitted yield.
    }
  \label{fig:bkinplots}
\end{figure}

\begin{figure}[!tbp]
  \centerline{\includegraphics[width=0.9\linewidth]{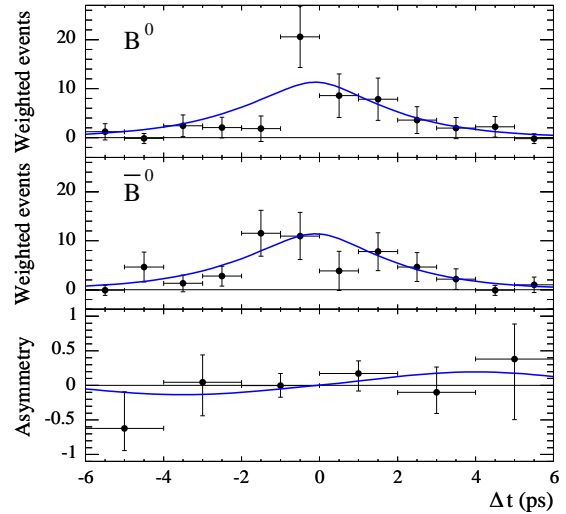}}
  \caption{Signal distribution for $\deltat$, obtained with the
    weighting technique described in the text, with $B_{\rm tag}$
    tagged as $\Bz$ (top) or $\Bzb$ (center), and the asymmetry ${\cal
      A}_{\KS\piz}(\deltat)$ (bottom). The curves represent the PDFs
    for signal decays in the likelihood fit.}
  \label{fig:dtplot}
\end{figure}

Figure~\ref{fig:bkinplots} shows the background-subtracted
distributions of \mb{} and \mmiss{} for all \Bztokspiz{} candidates in
the fit. The background subtraction is performed with an event
weighting technique~\cite{Pivk:2004ty}. Events contribute according to
a weight constructed from the covariance matrix for the yields ($N_S$
and $N_B$) and the probability ${\cal P}_S$ and ${\cal P}_B$ for the
event, computed without the use of the variable that is being
displayed. The curves represent the signal PDFs used in the fit. The
insets show the corresponding signal-subtracted distributions with the
background PDFs.  Figure~\ref{fig:dtplot} shows the
background-subtracted distributions of $\deltat$ for $\Bz$- and
$\Bzb$-tagged events, and of the asymmetry ${\cal
  A}_{\KS\piz}(\deltat) = \left[N_{\Bz} -
  N_{\Bzb}\right]/\left[N_{\Bz} + N_{\Bzb}\right]$ as a function of
$\deltat$.
  
The extraction of \deltat{} with the IP-constrained fit has been
extensively tested on large samples of simulated \Bztokspiz{} decays
with different values of $C$ and $S$.  We have also exploited a control sample
of approximately $1900$ observed \Bztojpsiks{} decays with
$\jpsi\to\mup\mun$ and $\jpsi\to\epem$, using the
procedure described in~\cite{Aubert:2004xf}.  Based on these studies
we assign a systematic uncertainty of $0.023$ on $S$ and $0.014$ on $C$
due to the \deltat{} reconstruction and the choice of the resolution
function.  As a cross-check we measure the $\Bz$ lifetime in
\Bztokspiz{} decays in data and find that it agrees with the world
average. We evaluate the effect of a possible misalignment of the SVT
by introducing misalignments in the simulation and assign a systematic
uncertainty of $0.020$ on $S$ and $0.007$ on $C$.  We also consider
large variations of the position and size of the interaction region,
which we find to have negligible impact. We include a systematic
uncertainty of $0.012$ on $S$ and $0.018$ on $C$ to account for
imperfect knowledge of the PDFs used in the fit.  Using simulated
events we estimate a contribution of $2.3\pm1.7$ events from other $B$
decays for which we assign a systematic uncertainty of $0.019$ on $S$
and $0.015$ on $C$.

The detection efficiency for signal events is obtained from a
Monte Carlo simulation. The efficiency of the \KS{}
selection is calibrated with a large sample of inclusive
$\KS\to\pip\pim$ decays. The $\piz\to\gamma\gamma$ efficiency is
calibrated with $\epem\to\taup\taum$ events with
$\tau^-\to\rhom\nu_{\tau}$. The systematic uncertainty associated with
the efficiency is \unit[$2.8$]{\%} for \KS{} and \unit[$3.0$]{\%} for
\piz{}. We assign additional systematic uncertainties of
\unit[$1.2$]{\%} for the $L_2/L_0$ cut, \unit[$2.0$]{\%} for the
selection on \mb{} and a total of \unit[$2.0$]{\%} for uncertainties
in the signal PDFs. Finally, we include a systematic uncertainty of
\unit[$1.4$]{\%} to account for unknown contributions from other $B$
decays and a systematic uncertainty of \unit[$0.6$]{\%} due to the
uncertainty in the total number of $\Y4S\to\BB$ decays.


In summary, we have reported improved measurements of the branching
fraction and \CP-violating asymmetry for the decay \Bztokspiz{}. The
measured values of \skspiz{} and \ckspiz{} are consistent with the
Standard Model predictions. The measured branching fraction is
consistent with measurements from other experiments~\cite{Brkspiz}.
These results supersede our previous measurements of the branching
fraction~\cite{Aubert:2003sg} and CPV asymmetries~\cite{Aubert:2004xf},
which were based on a subset of the data presented here.


\par
We are grateful for the excellent luminosity and machine conditions
provided by our \pep2\ colleagues, 
and for the substantial dedicated effort from
the computing organizations that support \babar.
The collaborating institutions wish to thank 
SLAC for its support and kind hospitality. 
This work is supported by
DOE
and NSF (USA),
NSERC (Canada),
IHEP (China),
CEA and
CNRS-IN2P3
(France),
BMBF and DFG
(Germany),
INFN (Italy),
FOM (The Netherlands),
NFR (Norway),
MIST (Russia), and
PPARC (United Kingdom). 
Individuals have received support from CONACyT (Mexico), A.~P.~Sloan Foundation, 
Research Corporation,
and Alexander von Humboldt Foundation.


\begin{thebibliography}{99}
\bibitem{ref:cc} Unless explicitly stated, conjugate decay modes are
assumed throughout this paper.
\end{thebibliography}
\end{document}